\documentclass[prb,aps,nobibnotes]{revtex4}
\usepackage{graphicx}%
\usepackage{dcolumn}
\usepackage{amsmath}
\usepackage{amssymb}
\usepackage{bm}
\begin{document}

\title{\centering\Large\bf Theory of solvation in polar nematics}
\author{Vitaly Kapko}
\author{Dmitry V.\ Matyushov}
\email[E-mail: ]{dmitrym@asu.edu.}

\date{\today}

\begin{abstract}
  We develop a linear response theory of solvation of ionic and
  dipolar solutes in anisotropic, axially symmetric polar solvents.
  The theory is applied to solvation in polar nematic liquid crystals.
  The formal theory constructs the solvation response function from
  projections of the solvent dipolar susceptibility on rotational
  invariants. These projections are obtained from Monte Carlo
  simulations of a fluid of dipolar spherocylinders which can exist
  both in the isotropic and nematic phase. Based on the properties of
  the solvent susceptibility from simulations and the formal solution,
  we have obtained a formula for the solvation free energy which
  incorporates experimentally available properties of nematics and the
  length of correlation between the dipoles in the liquid crystal. The
  theory provides a quantitative framework for analyzing the
  steady-state and time-resolved optical spectra and makes several
  experimentally testable predictions.  The equilibrium free energy of
  solvation is anisotropic in the nematic phase and is given by a quadratic function of
  cosine of the angle between the solute dipole and the solvent
  nematic director.  The sign of solvation anisotropy is determined by
  the sign of dielectric anisotropy of the solvent: solvation
  anisotropy is negative in solvents with positive dielectric
  anisotropy and \textit{vice versa}.  The solvation free energy is
  discontinuous at the point of isotropic-nematic phase transition.
  The amplitude of this discontinuity is strongly affected by the size
  of the solute becoming less pronounced for larger solutes. The
  discontinuity itself and the magnitude of the splitting of the
  solvation free energy in the nematic phase are mostly affected by
  microscopic dipolar correlations in the nematic solvent.
  Illustrative calculations are presented for the Stokes shift and
  Stokes shift correlation function of coumarin-153 in
  4-n-pentyl-4'-cyanobiphenyl (5CB) and 4,4-n-heptyl-cyanopiphenyl
  (7CB) solvents as a function of temperature in both the nematic and
  isotropic phase.
\end{abstract}

\affiliation{ Department of Chemistry and Biochemistry and the Center
  for the Study of Early Events in Photosynthesis, Arizona State
  University, PO Box 871604, Tempe, AZ 85287-1604} 

\date{\today }
\maketitle

\section{Introduction}
\label{sec:1}
The problem of polar solvation is one of the oldest problems of
Physical Chemistry which yet is still a field of active theoretical
and experimental research.  The calculation of solvation free energy
is particularly complex, since it is affected by a variety of
contributions including the short-range cavity formation energy,
medium-range dispersion and induction forces (non-polar solvation),
and long-range electrostatic interactions (polar solvation). These
components often compensate and complement each other when solvents of
different polarity are considered.  As a result, calculations of the
overall free energy of solvation are challenging and often require
phenomenological parametrization. Many phenomena encountered in
chemistry (spectroscopy, redox reactions, etc.)  are, however, less
affected by the cavitation free energy than by non-polar and polar
solvation, while the latter often dominates in polar solvents.
Therefore, much effort over the last 80 years, following the work of
Born,\cite{Born:20} Onsager,\cite{Onsager:36} and
Kirkwood\cite{Kirkwood:34} has focused on the understanding and
modeling of electrostatic, polar solvation.
  
The original Born-Onsager idea of calculating the electrostatic
solvation free energy as the continuum dielectric response to charges
of the solute has found broad applications, in particularly to
solvation of large molecules often encountered in bio-medical
research.\cite{Cramer:99,Tomasi:04} For smaller solutes, formal
liquid-state theories, most notably integral equation theories, have
found broad application. These theories are normally formulated either
in terms of site-site\cite{Raineri:99} or multipolar
interaction\cite{Richardi:98} potentials. The proliferation of
computer simulation techniques has helped to clarify many microscopic
features of solvation as well as to test and refine the formal models.

Most of the effort in the field of solvation thermodynamics has been
focused on the understanding of solvation in isotropic solvents.
Anisotropic solvation, important for chemical reactivity in biological
membranes, surfaces, and liquid crystalline solvents, has attracted
relatively little attention. Also, from the side of experiment, almost
nothing is known about thermodynamics of solvation in liquid crystals.
There is a very limited evidence on the solvatochromic shift from
spectroscopy\cite{Urisu:78} and a few solvation dynamics
studies\cite{Saielli:98,Bartolini:99,Rau:01} have been reported.  Computer experiment on
solvation in liquid crystals virtually does not exist. Continuum
models, representing the effect of solvent anisotropy by a tensorial
dielectric constant, have been
proposed.\cite{Mennucci:jcp:95,Mennucci:jpb:97} These approaches
provide a very useful continuum limit since, for instance, the Onsager
problem of solvation of a spherical dipole\cite{Onsager:36} has an
exact analytical solution for continuum
nematics.\cite{Urano:75,Urano:77}

Despite the progress in using dielectric continuum models, a few
fundamental problems still need to be resolved. First, limits of the
applicability of the continuum approximation to solvation in polar
nematics have not been established. Liquid crystals are mostly made of
bulky elongated molecules, and it is \textit{a priori} unclear if
continuum models can be applied to solvation of solutes of size often
comparable to the size of the solvent molecules. Second, it is not
clear if dielectric response of a liquid crystal to the solute electric
field can in principle be represented by a single quantity, the
dielectric constant, in particular close to the isotropic-nematic
phase transition.

The approach we propose in this paper is based on the recently
obtained microscopic solution for dipole solvation.\cite{DMjcp1:04}
The model is based on the assumption that the solute-solvent
interaction potential is given by the interaction of the solute
charges with the solvent dipolar polarization. The
solvation free energy is then expressed through the polarization autocorrelation
function of the pure solvent without any particular
assumptions regarding the solvent structure.  The theory is thus applicable to an
arbitrary \textit{isotropic} dielectric. The goal of this paper is to
generalize this approach to the case of a solvent with axial symmetry.
For solvation in isotropic liquids, two projections of the
polarization autocorrelation function, longitudinal and transverse,
are sufficient to describe the dipolar response. Lowering the symmetry
of the solvent requires a larger set of projections. We derive a
formally exact expression for the free energy of ionic and dipolar
solvation in Sec.\ \ref{sec:2} [Eq.\ (\ref{eq:2-46})].

The full formulation of the theory requires projections of the
polarization correlation function on rotational invariants. These are
obtained here from computer simulations of a fluid of dipolar hard
spherocylinders. The application of the theory to experiment requires,
however, a solution based on the input parameters available from
experiment. This formulation is given in Sec.\ \ref{sec:3} [Eq.\ 
(\ref{eq:3-9})] in form of a linear combination of solutions obtained
in the limit of zero wavenumber (continuum) and infinite wavenumber.
The relative contribution of each component depends on the correlation
length of dipolar fluctuations in the liquid (distinct from the
correlation length of the order parameter fluctuations in Landau-de
Gennes theory of liquid crystals \cite{deGennesLC}). One of the
principle results of this study is a very slow approach of the
solvation free energy to its continuum limit, thus invalidating
continuum approaches to solvation of small and medium-size solutes. We
study the dependence of the free energy of solvation on the angle
between the solute dipole and nematic director as well as on
temperature when crossing the point of isotropic-nematic transition in
Sec.\ \ref{sec:4}. The solvation free energy is shown to pass through
a discontinuity at the transition temperature and becomes anisotropic
in the nematic phase. Also, the Stokes shift correlation function
changes from a single-exponential decay in the isotropic phase to
bi-exponential decay in the nematic phase. Our results are summarized
in Sec.\ \ref{sec:5}.

\section{Theory}
\label{sec:2}
The linear response approximation (LRA) provides a solution for the
solvation free energy (strictly speaking, the chemical potential) $\mu$
in terms of the response function $\bm{\chi}(\mathbf{r}_1,\mathbf{r}_2)$
which gives the dipolar polarization in the point of space
$\mathbf{r}_1$ produced by the external electric field
$\mathbf{E}_0(\mathbf{r}_2)$ at the point of space $\mathbf{r}_2$:
\begin{equation}
  \label{eq:2-1}
   \mathbf{P}(\mathbf{r}_1) = \int \bm{\chi}(\mathbf{r}_1,\mathbf{r}_2)\cdot \mathbf{E}_0(\mathbf{r}_2) d\mathbf{r}_2 ,
\end{equation}
where subscript ``0'' for the variables refer to the solute. 
The solvation free energy is then
\begin{equation}
  \label{eq:2-2}
  \mu = - \frac{1}{2} \int\mathbf{P}(\mathbf{r}_1)\cdot\mathbf{E}_0(\mathbf{r}_1) d\mathbf{r}_1 .
\end{equation}
The dependence on two separate positions, instead of $\mathbf{r}_1 -
\mathbf{r}_2$ of homogeneous solvents, reflects the inhomogeneous
nature of the solvent response in the presence of the repulsive core
of the solute expelling the solvent from its volume. 

In $\mathbf{k}$-scape, Eq.\ (\ref{eq:2-2}) becomes
\begin{equation} 
\label{eq:2-3}
\mu = -\frac{1}{2} \int \frac{d \mathbf{k}_1 d \mathbf{k}_2}{(2 \pi)^6}
                   \tilde{\mathbf{E}}_0(\mathbf{k}_1) \cdot \pmb{\tilde \chi}(\mathbf{k}_1,
                    \mathbf{k}_2) \cdot \tilde{\mathbf{E}}_0(-\mathbf{k}_2) .
\end{equation}
Here, the Fourier transform of the electric field is taken over the solvent volume
$\Omega$ excluding the space occupied by the solute
\begin{equation}
  \label{eq:2-4}
  \tilde{\mathbf{E}}_0(\mathbf{k}_1) = \int_{\Omega} \mathbf{E}_0(\mathbf{r}) e^{i\mathbf{k}\cdot\mathbf{r}} d\mathbf{r} .
\end{equation}
The solute space is made by the van der Waals repulsive cores of its
atoms. The radii of the solute atoms exposed to the solvent are
augmented by the shortest distance to the solvent dipole, which, for
cylindrically symmetric molecules, is equal to the radius of the
cylindrical part of the molecule.  Further, the second-rank tensor
$\pmb{\tilde \chi}$ is
\begin{equation}
\label{eq:2-5}
\tilde{\chi}_{\alpha \beta}(\mathbf{k}_1,\mathbf{k}_2) = \frac{1}{k_B T} 
                        \langle \delta \tilde{P}_{\alpha}(\mathbf{k}_1) \delta \tilde{P}_{\beta}(\mathbf{-k}_2)\rangle_0 ,
\end{equation}
where $\delta \mathbf{\tilde P}(\mathbf{\tilde k})$ is the Fourier
transform of the fluctuation of the solvent dipolar polarization.
 
The LRA solution is independent of the electrostatic field of the
solute and the subscript ``0'' in the angular brackets denotes the
statistical average taken at the presence of a fictitious solute with
the repulsive core of the real solute but the electrostatic
solute-solvent coupling turned off.\cite{Chandler:93,DMjcp1:04} In a
hypothetical case of an infinitely small solute, $\pmb{\tilde \chi}$ is
equal to the dipolar susceptibility of pure solvent $\pmb{\tilde \chi}_s$
which depends on only one wavevector:
\begin{equation}
\label{eq:2-6}
\pmb{\tilde \chi}(\mathbf{k}_1,\mathbf{k}_2)= \delta_{\mathbf{k}_1,\mathbf{k}_2} 
\pmb{\tilde \chi}_s(\mathbf{k}_1) ,
\end{equation}
where subscript ``s'' denotes the solvent. 

In the general case, $\pmb{\tilde \chi}(\mathbf{k}_1,\mathbf{k}_2)$ is
affected by the presence of the solute and depends on two
$\mathbf{k}$-vectors. The effect of the solute on solvent response can
generally be separated into two major contribution. The repulsive core
of the solute distorts the local density of the solvent around it. The
spherically-symmetric solute-solvent pair correlation function
$h_{0s}(r)$ is then different from the solvent-solvent pair
correlation function $h_{ss}(r)$. This density disturbance affects the
dipolar polarization and, consequently, the response function.
Another, by far more significant, effect of the solute on the solvent
response function is related to the expulsion of the dipolar
polarization from the solute volume. In continuum models, this effect
is responsible for the surface charge at the dielectric cavity and,
when the cavity does not coincide with the equipotential surface,
results in a transverse component in the dielectric response. The
Maxwell's dielectric displacement\cite{Boettcher:73}
$\mathbf{D}(\mathbf{r})$ then differs from the external electric field
$\mathbf{E}_0(\mathbf{r})$.

The exclusion of the dipolar polarization from the solute volume is
accounted for in Chandler's Gaussian approximation
\cite{Chandler:93,Song:96} resulting in the following equation for the
$\mathbf{k}$-space response function\cite{DMjcp1:04}
\begin{equation}
\label{eq:2-7}
\pmb{\tilde \chi}(\mathbf{k}_1,\mathbf{k}_2) = \delta_{\mathbf{k}_1,\mathbf{k}_2}
                 \pmb{\tilde \chi}_s(\mathbf{k}_1) - \pmb{\tilde \chi}''(\mathbf{k}_1)\cdot  
\tilde{\theta}_0(\mathbf{k}_1-\mathbf{k}_2) \pmb{\tilde \chi}_s(\mathbf{k}_2) .
\end{equation}
Here, $\delta_{\mathbf{k}_1,\mathbf{k}_2} =
(2\pi)^3\delta(\mathbf{k}_1-\mathbf{k}_2)$ and $\tilde{\theta}_0(\mathbf{k})$ is the
Fourier transform of the step function, which equals to unity inside
the solute and is zero everywhere else. Further, in Eq.\ (\ref{eq:2-5}), 
\begin{equation}
\label{eq:2-8}
\pmb{\tilde \chi}''(\mathbf{k}) = \pmb{\tilde \chi}_s(\mathbf{k})\cdot  
\left[ \pmb{\tilde \chi}_s(\mathbf{k}) -  \pmb{\tilde \chi}'(\mathbf{k}) 
\right]^{-1}  ,
\end{equation}
where
\begin{equation}
  \label{eq:2-9}
   \pmb{\tilde \chi}'(\mathbf{k}) = \Omega^{-1} \int_{\Omega} d\mathbf{r_1} d\mathbf{r}_2 \pmb{\chi}_s(\mathbf{r}_1-\mathbf{r}_2)
    e^{i\mathbf{k}\cdot(\mathbf{r}_2-\mathbf{r}_1)} 
\end{equation}
and integration in Eq.\ (\ref{eq:2-9}) is over the volume $\Omega$ occupied by the solvent. 

The substitution of Eq.\ (\ref{eq:2-7}) into Eq.\ (\ref{eq:2-3}) results
in the chemical potential of solvation given by the sum of two components:
\begin{equation}
\label{eq:2-10}
\mu = \mu_{\text{h}} + \mu_{\text{corr}} .
\end{equation}
The first term, $\mu_{\text{h}}$, corresponds to the homogeneous response
approximation (subscript ``h'') which assumes that correlations of
dipolar polarization are not modified by the solute and
$\pmb{\chi}(\mathbf{k}_1,\mathbf{k}_2)$ can be approximated by dipolar
susceptibility of the pure solvent according to Eq.\ (\ref{eq:2-6}).
The only modification introduced by the solute is the cutoff of the
electric field inside the solute [Eq.\ (\ref{eq:2-4})]:
\begin{equation}
\label{eq:2-10-1}
- \mu_{\text{h}} =\frac{1}{2} \int \frac{d\mathbf{k}}{(2\pi)^3} \tilde{\mathbf{E}}_0(\mathbf{k}) 
\cdot \pmb{\tilde \chi}_s(\mathbf{k}) \cdot \tilde{\mathbf{E}}_0(-\mathbf{k}) .
\end{equation}  

\begin{figure}[htbp]
  \centering
  \includegraphics*[width=8cm]{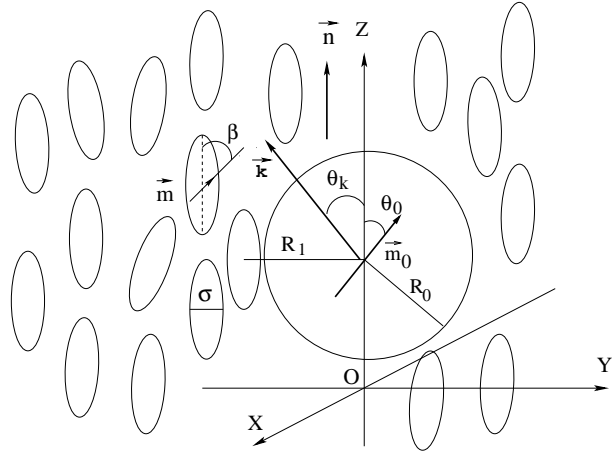}
  \caption{Dipolar solute in a nematic solvent. The laboratory system of coordinate
    is chosen to align the $z$-axis wit the nematic director
    $\mathbf{\hat n}$.  $\mathbf{m}_0$ denotes the direction of the
    solute dipole, $\mathbf{k}$ is the wave-vector. $\beta$ is the angle
    between the dipole moment and the long axis of the solvent
    molecule. }
  \label{fig:1}
\end{figure}

In isotropic solvents, the tensor $\pmb{\tilde \chi}_s$ is diagonal in a
coordinate system with one axis taken along $\bf{k}$. Its
eigenvalues, the longitudinal $\tilde \chi_s^L$ and transverse $\tilde
\chi_s^T$ projections, are quite different in the range $\sigma k < 2\pi$ due to
the long-range nature of the dipole-dipole interaction potential ($\sigma$
is the diameter of the solvent molecules).\cite{Hansen:03} In
particular, $\chi_s^T(0)$ grows as the dielectric constant $\epsilon$ in
strongly polar solvents, while $\chi_s^L(0)$ tends to a constant value.
Because of the mutual orthogonality of the longitudinal and transverse 
projections, the homogeneous solvation free energy $\mu_{\text{h}}$ splits into the
longitudinal (L) and transverse (T) parts, each given by a 3D integral
in $\mathbf{k}$:
\begin{equation}
  \label{eq:2-11}
  \mu_{\text{h}} = \mu_{\text{h}}^L + \mu_{\text{h}}^T, 
\end{equation}
where
\begin{equation}
  \label{eq:2-12}
  - \mu_{\text{h}}^{L,T} = \frac{1}{2} \int \frac{d\mathbf{k}}{(2\pi)^3} \chi_s^{L,T}(k)|\tilde E_0^{L,T}(\mathbf{k})|^2 .
\end{equation}
Once the transverse component of the field $\tilde E_0^T(\mathbf{k})$
is non-zero, which happens when either the solute shape or the solute
electric field deviates from spherical symmetry, $\mu_{\text{h}}^T$
grows linearly with the dielectric constant resulting in the
``transverse catastrophe''. Therefore, the homogeneous approximation
gives reasonable results only for spherical ions when the transverse
component of dipolar response is eliminated by the symmetry.  It is
the second term, generally represented by a 6D integral over the two
$\mathbf{k}$-vectors, that corrects (subscript ``corr'') for the
unphysical behavior of the transverse response.

The correction component $\mu_{\text{corr}}$ can be evaluated exactly
using analytical properties of the response function when the electric
field is known in the analytical form. In case of a dipole solute, the
final solution is conveniently formulated in terms of
$\mu_{\text{h}}^{L,T}$ (Ref.\ \onlinecite{DMjcp1:04}):
  \begin{equation}
    \label{eq:2-13}
    \mu = (\chi_{\text{tr}})^{-1}\left[ \chi_s^T(0)\mu_{\text{h}}^L  + \chi_s^L(0)\mu_{\text{h}}^T\right],
  \end{equation}
where
\begin{equation}
  \label{eq:2-14}
  \chi_{\text{tr}}=\frac{1}{3}\left(\chi_s^L(0) + 2 \chi_s^T(0)\right) .
\end{equation}

Unfortunately, this scheme is hard to implement for liquid crystals.
Since a liquid crystal has its own symmetry axis (Fig.\ \ref{fig:1}),
$\pmb{\tilde \chi}_s$ needs to be diagonalized for each value of
$\mathbf{k}$ making the problem rather complex.  Expansion of the solvent
response function in spherical harmonics \cite{Gubbins:84,Klapp:00}
appears to be a more straightforward way to the solution.  Our solution
below is given for the case of spherical ion (``i'') and spherical
dipole (``d'').  These two solutes are characterized by the following
expressions for the field Fourier transform in Eq.\ (\ref{eq:2-4})
\begin{equation}
\label{eq:2-17}
\tilde{\mathbf{E}}_0^{(i)}(\mathbf{k}) = 4\pi iq_0\frac{j_0(kR_1)}{k}\hat{\mathbf{k}}
\end{equation}
and 
\begin{equation}
\label{eq:2-18}
\tilde{\mathbf{E}}_0^{(d)}(\mathbf{k}) = -4\pi \frac{j_1(kR_1)}{kR_1} \mathbf{m}_0\cdot\mathbf{\hat{D}_k} .
\end{equation}
In Eqs.\ (\ref{eq:2-17}) and (\ref{eq:2-18}), $q_0$ and $\mathbf{m}_0$
are the solute charge and dipolar moment, and
$\mathbf{\hat{D}_k}=3\hat{\mathbf{k}}\hat{\mathbf{k}}-\mathbf{1}$ is
the dipolar tensor. Further, $R_0$ is the solute radius, $R_1=R_0+\sigma/2$
is the distance of closest solute-solvent separation, and
$\hat{\mathbf{k}}=\mathbf{k}/|\mathbf{k}|$.  Here we also use the
standard notation for the spherical Bessel functions $j_l(x)$.

The correction term in Eq.\ (\ref{eq:2-10}) is given by a double
$\mathbf{k}$-integral
\begin{equation}
\label{eq:2-15}
\mu_{\text{corr}}=\frac{1}{2} \int \frac{d\mathbf{k}_1 d\mathbf{k}_2}{(2\pi)^6} 
\tilde{\mathbf{E}}_0(\mathbf{k}_1) \cdot \pmb{\tilde \chi}''(\mathbf{k}_1)\cdot 
\tilde{\theta}_0(\mathbf{k}_1-\mathbf{k}_2) 
\pmb{\tilde \chi}_s(\mathbf{k}_2) \cdot \tilde{\mathbf{E}}_0(-\mathbf{k}_2) .
\end{equation} 
In order co convert it to a computationally tractable 3D integral, we
first introduce a direct-space field
\begin{equation}
\label{eq:2-16}
\mathbf{F}_0(\mathbf{r}) = \int \frac{d \mathbf{k}}{(2\pi)^3} e^{-i \mathbf{k} \cdot 
\mathbf{r}} \tilde{\mathbf{E}}_0(\mathbf{k})\cdot\pmb{\tilde \chi}''(\mathbf{k}) .
\end{equation}

Analytic properties of the response function $\pmb{\tilde\chi}''$ in
complex $k$-plane allows one to reduce the integration over
$\mathbf{k}$ to the angular integral over the directions of
$\mathbf{k}$.\cite{DMjcp1:04} For the field of a spherical dipole,
$\mathbf{F}_0=\mathbf{F}^{(d)}$ is constant within the solute
\begin{equation}
\label{eq:2-19}
\mathbf{F}^{(d)} = -\frac{1}{R_1^3} \int \frac{d \omega_k}{4 \pi}
      \mathbf{m}_0\cdot\mathbf{\hat{D}_k}\cdot\pmb{\tilde \chi}''(k=0) .
\end{equation}
For the spherical ion (see below), 
\begin{equation}
  \label{eq:2-20}
  \mathbf{F}^{(i)} = \mathbf{F}_0 = 0.   
\end{equation}
In order to calculate $\mathbf{F}_0$ in Eq.\ (\ref{eq:2-19}), we need
to obtain $\pmb{\tilde \chi}''(k=0)$ from the $k=0$ value of the solvent
dipolar susceptibility, which we consider next.

\subsection{Continuum limit}
\label{sec:2-1}
We use the laboratory Cartesian system of coordinates with $z$-axis
parallel to the nematic director $\hat{\mathbf{n}}$ (Fig.\ 
\ref{fig:1}).  The continuum limit for solvent response function can
be obtained from Maxwell's material equations with axially symmetric
dielectric constant characterized by longitudinal ($\epsilon_{\parallel}=\epsilon_z$) and
transverse ($\epsilon_{\perp}=\epsilon_x=\epsilon_y$) components:\cite{Caillol:pra:88}
\begin{equation}
\label{eq:2-21}
4 \pi \tilde{\chi}_{s,\alpha\beta}(k=0) = \left( \epsilon_{\alpha} -1 \right) \delta_{\alpha\beta} - \frac{\hat{k}_{\alpha}\hat{k}_{\beta} (\epsilon_{\alpha}-1)
      (\epsilon_{\beta}-1)}{\epsilon_{\perp}+(\epsilon_{\parallel}-\epsilon_{\perp}) (\hat{\mathbf{k}} \cdot \hat{\mathbf{n}})^2} .
\end{equation}
Note that $\tilde{\chi}_s(k=0)$ is an even function of
$\hat{\mathbf{k}}$; therefore, according to Eqs.\ (\ref{eq:2-17}) and
(\ref{eq:2-16}), $\mathbf{F}_0 = 0$ for the spherical ion [Eq.\ 
(\ref{eq:2-20})].

To proceed with the dipolar solute, we first calculate the integral
\begin{equation}
\label{eq:2-22}
A_{\alpha\beta} = - \int \frac{d \omega_k}{4\pi} \sum\limits_\gamma \hat{D}_{\mathbf{k},\alpha\gamma}\tilde{\chi}_{s,\gamma\beta}(k=0) .
\end{equation}
The matrix $\mathbf{A}$ is diagonal with the elements:
\begin{equation}
\label{eq:2-23}
A_{xx}=A_{yy}=\frac{\epsilon_{\perp}-1}{8\pi(\epsilon_{\parallel}-\epsilon_{\perp})}\left[ (2 \epsilon_{\parallel}-\epsilon_{\perp}+2) -\frac{\epsilon_{\parallel}}{\epsilon_{\perp}}(\epsilon_{\perp}+2) \psi \right] ,
\end{equation}
\begin{equation}
\label{eq:2-24}
A_{zz}=-\frac{\epsilon_{\parallel}-1}{4\pi(\epsilon_{\parallel}-\epsilon_{\perp})}\left[ \epsilon_{\perp}+2 -(\epsilon_{\parallel}+2) \psi \right],
\end{equation}
where
\begin{equation}
\label{eq:2-25}
\psi=\int\limits_0^1\frac{dz}{1+(\epsilon_{\parallel}/\epsilon_{\perp}-1)z^2}=\left\{\begin{array}{lcl} 
\arctan \sqrt{\epsilon_{\parallel}/\epsilon_{\perp}-1}/\sqrt{\epsilon_{\parallel}/\epsilon_{\perp}-1} &,& \epsilon_{\parallel}>\epsilon_{\perp} \\
\qquad 1 &,& \epsilon_{\parallel}=\epsilon_{\perp} \\
\ln{\left(\frac{1+\sqrt{1-\epsilon_{\parallel}/\epsilon_{\perp}}} {1-\sqrt{1-\epsilon_{\parallel}/\epsilon_{\perp}}}\right)}/
\left(2 \sqrt{1-\epsilon_{\parallel}/\epsilon_{\perp}} \right)  &,& \epsilon_{\parallel}<\epsilon_{\perp}. \\
\end{array} \right.
\end{equation}
Note that $A_{xx}$ and $A_{zz}$, as well as $\tilde{\chi}_{0,xx}$ and
$\tilde{\chi}_{0,zz}$ below, have no singularities at $\epsilon_{\parallel}=\epsilon_{\perp}$
because terms in the square brackets are proportional to $\epsilon_{\parallel}-\epsilon_{\perp}$
at $\epsilon_{\parallel}-\epsilon_{\perp} \ll 1$.

In the Appendix, we prove the relation
\begin{equation}
\label{eq:2-26}
\pmb{\tilde \chi}_0 \equiv
\pmb{\tilde \chi}_s(k=0)-\pmb{\tilde \chi}'(k=0) = \int\frac{d \omega_k}{4 \pi} \pmb{\tilde \chi}_s(k=0).
\end{equation}
Then $\pmb{\tilde \chi}_0$ is diagonal with the elements 
\begin{equation}
\label{eq:2-27}
\tilde{\chi}_{0,xx}=\tilde{\chi}_{0,yy}=\frac{\epsilon_{\parallel}-1} {8\pi(\epsilon_{\parallel}-\epsilon_{\perp})} \left[ (2\epsilon_{\parallel}-
\epsilon_{\perp}-1) -\frac{\epsilon_{\parallel}}{\epsilon_{\perp}} (\epsilon_{\perp}-1) \psi \right]
\end{equation}
and
\begin{equation}
\label{eq:2-28}
\tilde{\chi}_{0,zz}=- \frac{\epsilon_{\parallel}-1}{4\pi(\epsilon_{\parallel}- \epsilon_{\perp})} \left[ \epsilon_{\perp}-1 -(\epsilon_{\parallel}-1) \psi \right] .
\end{equation}
When Eqs.\ (\ref{eq:2-21}) and (\ref{eq:2-26}) are used in the
definition of $\pmb{\chi}''$ in Eq.\ (\ref{eq:2-8}), the final result for
the field $\mathbf{F}^{(d)}$ becomes
\begin{equation}
\label{eq:2-29}
   F_{\alpha}^{(d)} =\frac{m_{0,\alpha}}{R_1^3} R_{\alpha},
\end{equation}
where $\alpha$ stands for $x,y,z$ and
\begin{equation}
\label{eq:2-30}
       R_{\alpha}=\frac{\epsilon_{\alpha}-n_{\alpha}(\epsilon_{\alpha}+2)}{\epsilon_{\alpha}-n_{\alpha}(\epsilon_{\alpha}-1)} .
\end{equation}
In Eq.\ (\ref{eq:2-30}), the so-called depolarization factors are given by\cite{LandauLifshitz8}
\begin{equation}
\label{eq:2-31}
\begin{split}
      n_z&=\frac{\epsilon_{\parallel}(\psi-1)}{\epsilon_{\perp}-\epsilon_{\parallel}}, \\
      n_x&=n_y=\frac{1-n_z}{2} .
\end{split}
\end{equation}
In the isotropic limit, when $\epsilon_{\parallel}=\epsilon_{\perp}$, one gets $n_x=n_y=n_z=1/3$.

The field $\mathbf{F}_0$ defined by Eq.\ (\ref{eq:2-16}) is a
generalization of the reaction field, introduced by Onsager for a
point dipole,\cite{Onsager:36} to an arbitrary configuration of solute
charges in a solute of arbitrary shape. We have shown here that this
field reduces to expected limits in the case of spherical ionic and
dipolar solutes. In the former case, the reaction potential created by
the polar liquid within the cavity is constant, and the reaction field
is zero. In the latter case, the field is constant and our expression
in Eqs.\ (\ref{eq:2-29})--(\ref{eq:2-31}) coincides with the reaction
field in an axially anisotropic dielectric previously derived for a
spherical dipole by solving the Poisson equation.\cite{Urano:77}
 
The zero reaction field in the case of a spherical ion eliminates the
correction term in Eq.\ (\ref{eq:2-10}). This means that the solvent
response is longitudinal and the dielectric displacement $\mathbf{D}$
is equal to the external field $\mathbf{E}_0$.  The free energy of
solvation is then fully determined by the homogeneous solvation term
[Eq.\ (\ref{eq:2-10-1})].  In case of a dipole, the solvent response
includes a transverse component, the dielectric displacement is not
equal to the external field, and the correction term is necessary:
\begin{equation}
\label{eq:2-33}
\mu_{\text{corr}}=\frac{1}{2 R_1^3} \sum\limits_{\alpha\beta} R_{\alpha} m_{0,\alpha}\,
                   \int \frac{d \mathbf{k}}{(2\pi)^3} \tilde{\chi}_{s,\alpha\beta}(\mathbf{k})\,
                      \tilde{\theta}_0(\mathbf{k})\, \tilde{E}_{0,\beta}(-\mathbf{k}),
\end{equation}
where for a spherical solute 
\begin{equation}
  \label{eq:2-34}
   \tilde{\theta}_0(k)=4\pi R_1^3 \frac{j_1(kR_1)}{kR_1}.  
\end{equation}

Equations (\ref{eq:2-10-1}) and (\ref{eq:2-33}) give the correct
continuum limit (subscript ``c'') for the solvation free energy after
the replacement of $\pmb{\tilde \chi}_s(k)$ with its value at $k=0$:
\begin{equation}
\label{eq:2-35}
\mu_{\text{c}}^{(i)}  = - \frac{q^2}{2R_1}\left( 1 -\frac{\psi}{\epsilon_{\perp}} \right)
\end{equation}
for the ion and
\begin{equation}
\label{eq:2-36}
\mu_{\text{c}}^{(d)} = - \frac{1}{2} \mathbf{m}_0 \cdot \mathbf{F}_0^{(d)} =
- \frac{m_0^2}{2\, R_1^3} \left[ R_{x} + (R_{z}-R_{x})\cos^2\theta_0 \right]
\end{equation}
for the dipole. In Eq.\ (\ref{eq:2-36}), $\theta_0$ is the angle between
the solute dipolar moment and the director (Fig.\ \ref{fig:1}).  In
the limit $\epsilon_{\parallel}\to\epsilon_{\perp}$, Eqs.\ (\ref{eq:2-35}) and (\ref{eq:2-36})
reduce to their well-know isotropic counterparts, the Born
formula\cite{Born:20}
\begin{equation}
  \label{eq:2-37}
  \mu_{\text{B}}^{(i)} = - \frac{q^2}{2R_1}\left( 1 -\frac{1}{\epsilon} \right)
\end{equation}
and the Onsager formula\cite{Onsager:36}
\begin{equation}
  \label{eq:2-38}
  \mu_{\text{O}}^{(d)}=-\frac{m_0^2}{R_1^3}\,\frac{\epsilon-1}{2\epsilon+1}.
\end{equation}
Note that the cavity radius is not specified in continuum models.
However, empirical experience suggests using the van der Waals radius
$R_0$ in place of the radius of closest solute-solvent approach $R_1$
appearing in microscopic solvation models.

\subsection{Microscopic theory} 
\label{sec:2-2}
The dependence on the orientation of the wave-vector in Eqs.\ 
(\ref{eq:2-10-1}) and (\ref{eq:2-33}) can be integrated out by expanding
the solvent dipolar susceptibility $\pmb{\tilde \chi}_s$ in spherical
harmonics [Eq.\ (\ref{A4})]:
\begin{equation}
\label{eq:2-39}
\tilde{\chi}_{s,n_1 n_2} (\mathbf{k}) = \sum\limits_l \tilde{\chi}_{s,n_1 n_2l}(k) Y^*_{l,-n_1-n_2}(\omega_k),
\end{equation}
where $\omega_k$ denotes the orientation of $\mathbf{\hat k}$.  The
spherical components of the solute electric field can be obtained for
the ionic and dipolar solutes:\cite{Gubbins:84}
\begin{equation}
\label{eq:2-40}
\tilde{E}^{(i)}_{0,n_1}(\mathbf{k}) = 4\pi\sqrt{\frac{4\pi}{3}}q_0 i \frac{j_0(kR_1)}{k}Y^*_{1,n_1}(\omega_k)
\end{equation}
and 
\begin{equation}
\label{eq:2-41}
\tilde{E}^{(d)}_{0,n_1}(\mathbf{k}) = - 16\pi^2 \sqrt{\frac{2}{5}} m_0 \frac{j_1(kR_1)}{kR_1} 
\sum\limits_{n_1'} C(112; n_1, n_1', n_1+n_1') Y_{1,n_1'}(\hat{m}_0) Y^*_{2,n_1+n_1'}(\omega_k) .
\end{equation}
In Eq.\ (\ref{eq:2-41}), $C(l_1 l_2 l; n_1, n_2, n)$ are the Clebsch-Gordan coefficients.\cite{Gubbins:84}
Using the product rule 
and orthogonality of spherical harmonics\cite{Gubbins:84} we obtain
\begin{equation}
\label{eq:2-42}
\begin{split}
\mu_h^{(i)}&=-4q_0^2\sum\limits_{n_1n_2l}\int\limits_0^{\infty}dk\,j_0^2(kR_1) \chi_{s,n_1n_2l}(k)(-1)^{n_1+n_2}
                            \frac{C(11l;0,0,0)}{2l+1}C(11l;n_1,n_2,n_1+n_2),\\
\mu_h^{(d)}& = - \frac{24 m_0^2}{R_1^2} \int\limits_0^{\infty} dk j_1^2(kR_1) 
\sum\limits_{n1 n_2 l} \chi_{s, n_1 n_2 l}(k) (-1)^{n_1+n_2} \frac{C(22l;000)}{2l+1}\\
&\sum\limits_{n'} C(112;n_1, n', n_1+n') C(112;n_2, -n', n_2-n') C(22l;n_1+n',n_2-n',n_1+n_2) \hat{m}_{0,n'} \hat{m}_{0,-n'},\\
\mu_{\text{corr}}^{(d)} & = -\frac{4\sqrt{6} m_0^2}{5 R_1^2} \sum\limits_{n_1 n_2} \int\limits_0^{\infty} dk
j_1^2(kR_1) \tilde{\chi}_{s,n_1 n_2 2}(k) (-1)^{n_2} C(112;n_1, n_2, n_1+n_2)
\hat{m}_{0,n_1} \left( \hat{m}_0 R \right)_{-n_1},\\
\mu_{\text{corr}}^{(i)} &=0 .
\end{split}
\end{equation}
In Eq.\ (\ref{eq:2-42}),
\begin{equation}
\label{eq:2-43}
    \chi_{s, n_1 n_2 l}(k) = \sqrt{\frac{2l+1}{4\pi}} \tilde{\chi}_{s,n_1 n_2 l}(k),
\end{equation}
and the relations between the spherical and Cartesian components of the vector 
$\mathbf{\hat m}=\mathbf{m}/m$ are 
\begin{equation}
  \label{eq:2-44}
  \begin{split}
    \hat{m}_{0,0} &= \hat{m}_{0,z},\\
    \hat{m}_{0,1} &=-(\hat{m}_{0,x}+i \hat{m}_{0,y})/\sqrt{2},\\
    \hat{m}_{0,-1}&=(\hat{m}_{0,x}-i \hat{m}_{0,y})/\sqrt{2}.
  \end{split}
\end{equation}
Similarly, for the second-rank tensor, one has
\begin{equation}
  \label{eq:2-45}
 \begin{split}
  (\hat{m}_0R)_0   & = \hat{m}_{0,z} R_{z}, \\
  (\hat{m}_0R)_1   & =-(\hat{m}_{0,x}+i \hat{m}_{0,y})R_{x}/\sqrt{2}, \\
  (\hat{m}_0R)_{-1}  & =(\hat{m}_{0,x}-i \hat{m}_{0,y})R_{x}/\sqrt{2}.  
 \end{split}
\end{equation}
Similar relations exist for the components of the second-rank tensor
$\pmb{\tilde \chi}_s(\mathbf{k})$.\cite{Gubbins:84}

For weakly polar solvents the Cartesian components of $\pmb{\tilde
  \chi}_s(\mathbf{k})$ form a diagonal matrix in the laboratory
system of coordinates specified in Fig.\ \ref{fig:1}.  Moreover, the
components related to axes $x$ and $y$ are almost equal to each
others.  In this approximation the solvation free energy reduces to
\begin{equation}
\label{eq:2-46}
\begin{split}
\mu_{\text{h}}^{(i)} & =-\frac{4q_0^2}{3} \int\limits_0^{\infty}\, dk\, j_0^2(kR_1) A(k),\\
\mu_{\text{h}}^{(d)} & =-\frac{4 m_0^2}{5R_1^2} \int\limits_0^{\infty} dk j_1^2(kR_1) \left[ B(k)+C(k) \cos^2\theta_0 \right],\\
\mu_{\text{corr}}^{(d)} &= \frac{4 m_0^2}{5 R_1^2} \int\limits_0^{\infty} dk j_1^2(kR_1) \left[ R_{x} \chi_{s,xx2}(k) -\left( 2 R_{z} \chi_{s,zz2}(k) +
R_{x} \chi_{s,xx2}(k) \right) \cos^2\theta_0 \right], \\
\mu_{\text{corr}}^{(i)} & =0,
\end{split}
\end{equation}
where 
\begin{equation}
\label{eq:2-47}
\begin{split}
A(k) & =2\chi_{s,xx0}(k)+\chi_{s,zz0}(k)+\frac{2}{5}\left(\chi_{s,zz2}(k)-\chi_{s,xx2}(k)\right),\\
B(k) & =3\chi_{s,zz0}(k)+7\chi_{s,xx0}(k)+\frac{1}{7}\left(3\chi_{s,zz2}(k)-10\chi_{s,xx2}(k)\right) -
\frac{4}{7} \left(\chi_{s,zz4}(k)-\chi_{s,xx4}(k) \right),\\
C(k) & =\chi_{s,zz0}(k)-\chi_{s,xx0}(k)+\frac{1}{7}\left(5\chi_{s,zz2}(k)+16\chi_{s,xx2}(k)\right)+
\frac{12}{7}\left(\chi_{s,zz4}(k)-\chi_{s,xx4}(k)\right).
\end{split}
\end{equation}
In Eqs.\ (\ref{eq:2-46}) and (\ref{eq:2-47}), $\chi_{s,\alpha\alpha l}(k)$ are the coefficients of expansion of 
the solvent response function in Legendre polynomials $P_l(\cos \theta_k)$:
\begin{equation}
\label{eq:2-48}
\tilde{\chi}_{s,\alpha\alpha}(\mathbf{k}) = \sum\limits_l \chi_{s,\alpha\alpha l}(k) P_l(\cos \theta_k).
\end{equation}

\section{Results}
\label{sec:3}
The theory developed in the previous section requires static
dielectric constants and dipolar susceptibility of the nematic solvent
as input.  Here we obtain these data from Monte Carlo (MC) simulation
of hard spherocylinders with embedded point dipoles.\cite{DMjcp2:03}
This fluid transforms from isotropic to nematic phase with decreasing
density.\cite{McGrother:98} NVT MC simulations of $N=800$ hard
spherocylinders were carried out in our previous
study.\cite{DMjcp2:03} The dipole moment $\mathbf{m}$ is parallel to
the cylinder axis and the aspect ratio of the length $L$ of the
cylindrical part of the molecule to its diameter $\sigma$ is equal to 5.
The thermodynamic state of this fluid is fully defined by two
parameters: the reduced dipole moment $(m^*)^2=m^2/(k_BT\sigma^3)$ and the
packing fraction $\eta=(\pi/6)\rho\sigma^3(1+3L/2\sigma)$, where $\rho=N/V$ is the solvent
number density.  Details of the simulation protocol are given in Ref.\ 
\onlinecite{DMjcp2:03}.

Simulations of nematics with high magnitudes of the dipole moment
$m^*$ are hindered by the tendency of neighboring dipoles to orient in
a locally antiferroelectric order.  Combined with the elongated shape
of the spherocylinders, local antiparallel alignment of dipoles
creates bottlenecks in the system phase space, which are hard to
explore by standard simulation techniques.\cite{McGrother:98} In
addition, the fluid of dipolar spherocylinders becomes smectic at
$(m^*)^2 > 6.0$ and $\eta =0.47$. Because of the relatively high aspect
ratio of the solvent molecules, the dipole moment in the range $0\leq
(m^*)^2 \leq 6.0$ gives a relatively small overall density of dipoles
and, therefore, a low dielectric constant. As a result, the
capabilities of simple hard-core models are rather limited in
exploring high-polarity nematics. The structure of real polar nematic
liquids, which can demonstrate rather high dielectric
constants,\cite{LC:01} is mostly determined by dispersion site-site
interactions between elongated molecules.

\begin{figure}[htbp]
  \centering
  \includegraphics*[width=8cm]{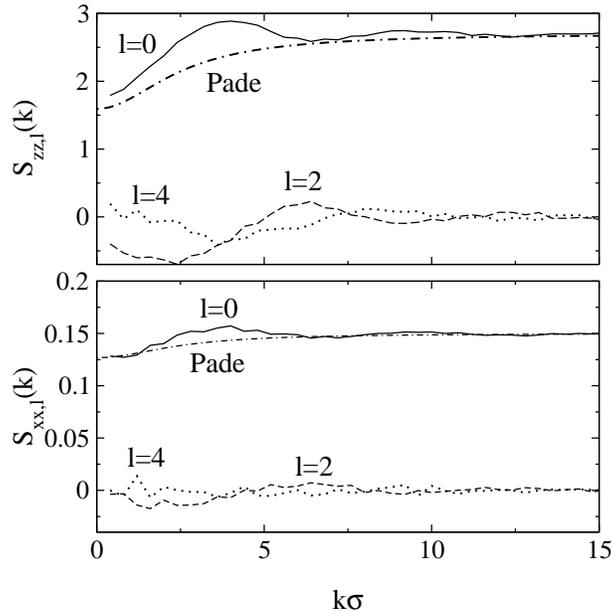}
  \caption{$S_{zz,l}(k)$ (upper panel) and $S_{xx,l}(k)$ (lower panel) projections of 
    dipolar structure factors for the fluid of hard spherocylinders:
    $l=0$ (solid line), $l=2$ (dashed line) and $l=4$ (dotted line).
    The dash-dotted lines refer to the Pad\'e  approximation
    [Eq.\ (\ref{eq:3-9})] with $\Lambda = 0.4 \sigma$. The structure factors are obtained
    from NVT MC simulations\cite{DMjcp2:03} of 800 solvent molecules
    with the packing density $\eta = 0.47$ and the dipole moment
    $m^2/(k_{\text{B}}T\sigma^3)  = 1.0$. The fluid is in the nematic
    phase with the nematic order parameter $S_2=0.85$. }
  \label{fig:2}
\end{figure}

For weakly polar nematics, the tensor of the dipolar susceptibility
$\pmb{\tilde \chi}_s$ is nearly diagonal in the Cartesian coordinate system
specified in Fig.\ \ref{fig:1}, with almost equal $xx$ and $yy$
projections.  Therefore, the formally exact formulas in Eq.\ (\ref{eq:2-42})
can be replaced by the approximate relation in Eq.\ (\ref{eq:2-46}).
The solvation free energy then depends on six one-dimensional
projection $\chi_{s,\alpha\alpha l}(k)$ [Eq.\ (\ref{eq:2-48})]. The corresponding
structure factors of dipolar polarization
\begin{equation}
  \label{eq:3-1}
  S_{\alpha\alpha,l}(k)=(4\pi/3y)\chi_{s,\alpha\beta l}(k)
\end{equation}
have been obtained here from equilibrium MC configurations of the fluid
of dipolar spherocylinders (Fig.\ \ref{fig:2}).\cite{DMjcp2:03}
In Eq.\ (\ref{eq:3-1}),
\begin{equation}
  \label{eq:3-2}
  y=4\pi m^2\rho/(9k_BT)
\end{equation}
is the standard dipole density parameter of dielectric
theories.\cite{Boettcher:73} The reduced dipole moment and the packing
fraction of the system have values $(m^*)^2=1$ and $\eta=0.47$ (the
isotropic-nematic phase transition occurs at $\eta_{\text{IN}}\approx0.407$).  In this
thermodynamic state, the nematic order parameter $S_2$ is equals to
$0.85$, and the longitudinal ($\epsilon_{\parallel}$, parallel to the director) and
transverse ($\epsilon_{\perp}$, perpendicular to the director) dielectric
constants are $1.89$ and $1.06$, respectively.

The noise in the structure factors obtained from simulations (Fig.\ 
\ref{fig:2}) arises from the fluctuations of the nematic director in
the laboratory system of coordinates attached to the simulation box.
This setup is necessary to insure that the wavevectors used to
calculate $S_{\alpha\alpha,l}(k)$ are eigenvectors of the periodic replicas of
the cubic simulation cell.  The calculations show that the
longitudinal structure factors ($S_{zz,l}$) are significantly larger
than the transverse structure factors ($S_{xx,l}$), as expected for nematics with
longitudinal dipolar moment. In addition, the magnitudes of
projections decrease rapidly with increasing index $l$.

Exact analytical expressions are available for the polarization
structure factors in $k\to0$ and $k\to\infty$ limits.  The continuum limit
relates $S_{zz,l}(0)$ and $S_{xx,l}(0)$ to the anisotropic static
dielectric constants through Eqs.\ (\ref{eq:2-21}), (\ref{eq:2-48}),
and (\ref{eq:3-1}). To calculate the $k\to\infty$ limit, we note that the
Cartesian components of the dipolar structure factors are given by the
following expression
\begin{equation}
\label{eq:3-3}
S_{\alpha\beta}(\mathbf{k})=(3/N)\sum\limits_{ij} \langle \hat{m}_{i,\alpha} \hat{m}_{j,\beta} 
                 e^{-i \mathbf{k}\cdot\mathbf{r}_{ij}} \rangle ,
\end{equation}
where $N$ is a number of solvent particles. All terms in Eq.\ (\ref{eq:3-3})
with $r_{ij}\neq0$ vanish at $k\to\infty$ resulting in
\begin{equation}
\label{eq:3-4}
       S_{\alpha\beta}(k\to\infty)=(3/N)\sum\limits_{i} \langle \hat{m}_{i,\alpha} \hat{m}_{i,\beta} \rangle =3\langle\hat{m}_{\alpha} \hat{m}_{\beta}\rangle .
\end{equation}
This yields
\begin{equation}
\label{eq:3-5}
\begin{split}
             S_{xx}(k\to\infty) & = S_{yy}(k\to\infty)=1-S_2P_2(\cos\beta),\\
             S_{zz}(k\to\infty) & = 1+2S_2P_2(\cos\beta),\\
             S_{zx}(k\to\infty) & = S_{zy}(k\to\infty)= S_{xy}(k\to\infty) =0 ,
\end{split}
\end{equation}
where $\beta$ is the angle between the dipole moment and the long
molecular axis of the solvent molecule (Fig.\ \ref{fig:1}) and
$P_2(x)$ is the second Legendre polynomial.

From Eqs.\ (\ref{eq:2-46}), (\ref{eq:2-47}), (\ref{eq:3-1}),
(\ref{eq:3-4}), and (\ref{eq:3-5}), we find the expression for the
solvation free energy when the $k\to\infty$ limit is used for the solvent
susceptibility:
\begin{equation}
   \label{eq:3-6}
         \mu_{\infty}^{(i)}=-3y\frac{q_0^2}{2R_1}
\end{equation}
and
\begin{equation}
     \label{eq:3-7}
             \mu_{\infty}^{(d)}=-\frac{m_0^2y}{R_1^3}\left( 1 + \frac{1}{5}S_2P_2(\cos\beta)P_2(\cos\theta_0)\right) .
\end{equation}

The continuum [$k=0$, Eqs.\ (\ref{eq:2-35}) and (\ref{eq:2-36})] and
short wave-length [$k\to\infty$, Eqs.\ (\ref{eq:3-6}) and (\ref{eq:3-7})]
limits are two asymptotes for the solvation free energy obtained by
setting, respectively, the constant $\pmb{\chi}_s(0)$ and $\pmb{\chi}_s(\infty)$
values for the solvent susceptibility in the $k$-integrals in Eqs.\ 
(\ref{eq:2-10-1}) and (\ref{eq:2-33}).  The advantage of these
limiting expressions is their simplicity and direct connection to
experimentally available properties of liquid crystalline solvents.
These limits can be used to derive a practically useful analytical
formula for $\mu$. The largest contribution to $\mu$ comes from the region
of $k$ where the squared spherical Bessel functions $j_l(kR_1)$ in
Eq.\ (\ref{eq:2-46}) has a maximum. This is the region around $k=0$
for the ionic solute and $k\approx2/R_1$ for the dipolar solute.  We need,
therefore, a continuous approximation for the structure factors that
generates a weighted linear combination of $k=0$ and $k\to \infty$ limits for
the solvation free energy.

\begin{figure}[htbp]
  \centering \includegraphics*[width=8cm]{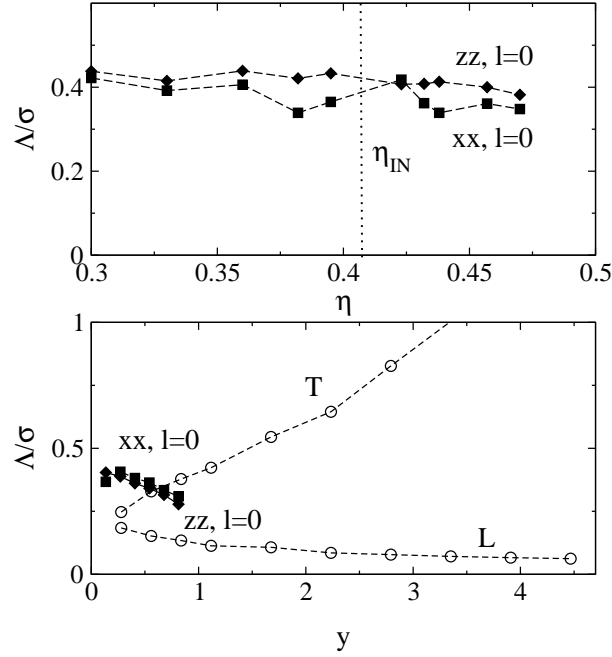}
  \caption{The length of dipolar correlations in the fluid
    of hard spherocylinders vs the solvent packing density (upper
    panel, $(m^*)^2=1.0$) and vs dipolar density $y$ (lower panel,
    $\eta=0.47$).  The data are obtained from NVT MC simulations. Closed
    diamonds refer to $zz, l=0$ projection, closed squares indicate
    $xx, l=0$ projection. Open points in the lower panel indicate the
    transverse and longitudinal correlation length obtained for a
    fluid of dipolar hard spheres with constant density $\rho\sigma^3=0.8$ and
    changing dipole moment.  The dashed lines connect the simulation
    points. The dotted vertical tine in the upper panel indicates the
    density of the isotropic-nematic phase transition,
    $\eta_{\text{IN}}=0.407$.  }
  \label{fig:3}
\end{figure}

The projections of the solvent susceptibility on spherical harmonic
are smooth functions of the wavevector for weakly polar nematics.  The
Cartesian components of the dipolar structure factors can, therefore,
be reasonably well approximated by Pad\'e forms interpolating between
the $k=0$ and $k\to \infty$ limits
\begin{equation}
  \label{eq:3-8}
       S_{\alpha\beta,l}(k)=\frac{S_{\alpha\beta,l}(0)+S_{\alpha\beta,l}(\infty) \Lambda_l^2k^2}{1+\Lambda_l^2k^2} .
\end{equation}
This formula introduces a new theory parameter, the polarization
correlation length $\Lambda_l$.  The correlation length can be extracted
from structure factors obtained from computer simulations by fitting
the slope of $S_{\alpha\beta,l}(k)$ vs $k^2$ to the $k\to 0$ expansion of Eq.\ 
(\ref{eq:3-8}) ($k< k_{max}\simeq 2/ \sigma $). Values of $\Lambda_l$ for the fluid of
dipolar hard spherocylinders depending on packing fraction $\eta$ and
dipolar density $y$ are shown in Fig.\ \ref{fig:3} (closed points).
$\Lambda_l$ was found to be rather weakly dependent on $\eta$, even through the
isotropic-nematic phase transition (Fig.\ \ref{fig:3}, upper panel).
The dependence on $y$ was obtained at fixed packing fraction $\eta=0.432$
and $(m^*)^2$ changing from $1.0$ to $6.0$.

We compare these results to the correlation lengths extracted from
slopes of the longitudinal structure factor $S^L(k)$ vs $k^2$ and the
inverse structure factor $1/S^T(k)$ vs $k^2$.  The latter definition
corresponds to the Ornstein-Zernike-Debye plot used for the scattering
function of liquids close to the critical temperature when the
structure factor is a decaying function of $k$.\cite{Stanley:87} On
the contrary, the longitudinal structure factor is a rising function
of $k$ requiring the direct expansion of $S^L(k)$ in $k^2$. The
correlation lengths for longitudinal and transverse dipolar
fluctuations in isotropic liquids are substantially different. This is
because these two projections mix together harmonics of the pair
distribution function with different index $l$
\begin{equation}
\begin{split}
S^{L}(k) & = 1 + \frac{\rho}{3}\left( \tilde h^{110}(k) + 2 \tilde h^{112}(k) \right),\\
S^{T}(k) & = 1 + \frac{\rho}{3}\left( \tilde h^{110}(k) - \tilde h^{112}(k) \right) . 
\end{split}
\end{equation}
where $\tilde h^{lmn}(k)$ is the Hankel transform [Eq.\ (\ref{A2})]. 

\begin{figure}[htbp]
  \centering
  \includegraphics*[width=8cm]{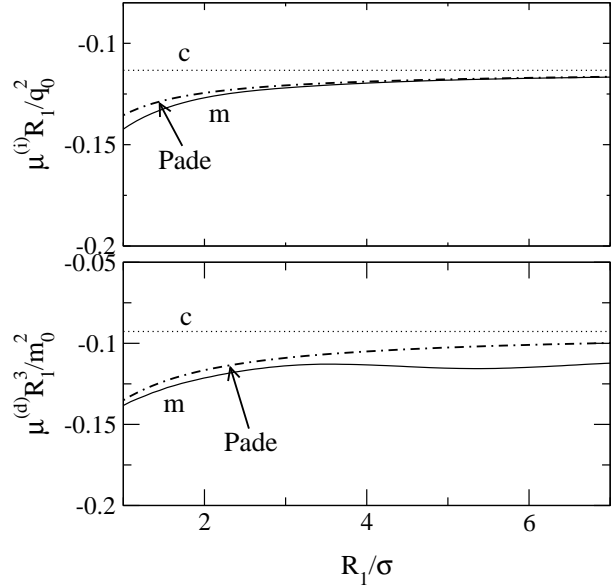}
  \caption{Solvation free energy  of ionic (upper panel, in $q_0^2/R_1$ units) and 
    dipolar (lower panel, in $m_0^2/R_1^3$ units) solutes vs $R_1$.
    Shown are the microscopic calculation [``m'', Eqs.\ 
    (\ref{eq:2-46})--(\ref{eq:2-48})], the Pad\'e form [``Pade'', Eq.\ 
    (\ref{eq:3-9})], and the continuum limit [``c'', Eqs.\ 
    (\ref{eq:2-35}) and (\ref{eq:2-36})]. }
  \label{fig:5}
\end{figure}

In contrast to longitudinal and transverse projections of isotropic
fluids, the projections $S_{\alpha\beta,l}(k)$ correspond to the same index
$l$. The anisotropy of the nematic phase relative to the director is
then taken out to the Legendre polynomial $P_l(\cos \theta_k)$ [Eq.\ 
(\ref{eq:2-48})]. In addition, we found that $\Lambda_l$ calculated from
different harmonics with the same $l$ are approximately equal to each
other, at least for weak polar solvents. This is why $\Lambda_l$ in Eq.\ 
(\ref{eq:3-8}) does not include Cartesian projections $\alpha,\beta$.  The
results for $l=0$ are presented in Fig.\ \ref{fig:5}, while data for
$l>0$ do not converge well because of large statistical errors.

\begin{figure}[htbp]
  \centering
  \includegraphics*[width=8cm]{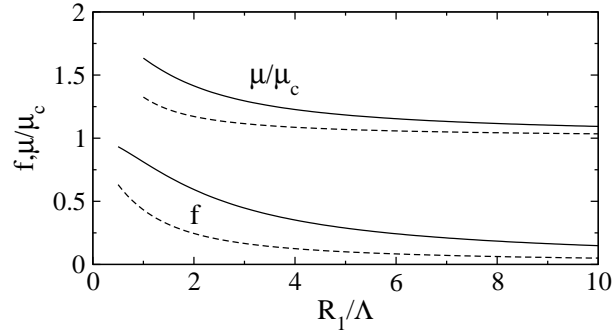}
  \caption{Non-locality functions [Eq.\ (\ref{eq:3-10})] and reduced solvation 
    free energy $\mu/ \mu_c$ vs the solute size.  The dashed and solid
    lines refer to the spherical ion and spherical dipole solutes,
    respectively.  }
  \label{fig:4}
\end{figure}

With the structure factors given by Eq.\ (\ref{eq:3-8}), the solvation
free energy is a linear combination of continuum and large wavevector
limits:
\begin{equation}
   \label{eq:3-9}
        \mu=\mu_c+f(R_1/\Lambda)\left(\mu_{\infty}-\mu_c\right) .
\end{equation}
The function $f(R_1/\Lambda)$ represents the contribution of the non-local
solvent response, influenced by the finite length of dipolar
correlations, to the solvation thermodynamics.  For the cases of
spherical ionic (i) and dipolar (d) solutes, this function is given by
the following expressions
\begin{equation}
\label{eq:3-10}
\begin{split}
   f^{(i)}(x) & = 0.5 \left[ 1 - e^{-2x} \right]/x,\\
   f^{(d)}(x) & = 1.5 \left[ x^2-1 + \left(x+1\right)^2e^{-2x} \right]/x^3 .
\end{split}
\end{equation}

Non-locality functions $f^{(i)}(R_1/ \Lambda)$ and $f^{(d)}(R_1/ \Lambda)$ decay
monotonically from one to zero with increasing the solute size $R_1$
(Fig.\ \ref{fig:4}).  The ionic non-locality function decays faster
than the dipolar function indicating that ionic solvation is better
described by continuum approximation than is dipolar solvation.
Notice that formula (\ref{eq:3-9}) applies to solvation in strongly
polar nematics, because it has been derived without assuming weak
polarity of the solvent (the cross-terms omitted in Eq.\ 
(\ref{eq:2-46}) are included in $\mu_{\text{c}}$).  Figure \ref{fig:5}
demonstrates a satisfactory agreement between the approximate solution
given by Eq.\ (\ref{eq:3-9}) and the exact microscopic theory [Eqs.\ 
(\ref{eq:2-46})--(\ref{eq:2-48})].  The increase of the solvent dipole
moment leads to an oscillatory character of the response function, but
even in this case, the Pad{\'e} approximation [Eq.\ (\ref{eq:3-8})]
gives reasonably accurate estimates for the solvation free energy in
isotropic polar solvents.\cite{DMmp:93}

Equation (\ref{eq:3-9}), combining the limiting continuum and high
wavevector values for ionic [Eqs.\ (\ref{eq:2-35}) and (\ref{eq:3-6})]
and dipolar [Eqs.\ (\ref{eq:2-36}) and (\ref{eq:3-7})] solutes, is the
central results of this study. The solvation free energy requires the
following solvent parameters: dielectric constants $\epsilon_{\parallel}$ and
$\epsilon_{\perp}$, the order parameter $S_2$, the polarization correlation
length $\Lambda$. In addition, the ionic charge or dipole moment, along with
the radius $R_1=R_0 + \sigma /2$, should be supplied. The dielectric
constants and the order parameter come from experiment. The
correlation length $\Lambda$ is not experimentally available and, for the
sake of interpreting the experiment, is, at the moment, a theory
parameter requiring fitting to some experimental observable.
Simulations of the model fluid of hard spherocylinders suggest
magnitudes of $\Lambda$ of the order $\Lambda\simeq 0.3 - 0.4 \sigma$.

\section{Theory predictions}
\label{sec:4}
The present theory allows us to make some specific predictions
regarding equilibrium solvation and solvation dynamics. The
electrostatic component of solvation free energy can be measured from
the steady-state Stokes shift of optical lines, whereas solvation dynamics
is probed by the Stokes shift correlation function. Our discussion
below will therefore target these two properties.

\subsection{Equilibrium solvation}
\label{sec:4-1}
Our present development highlights several issues important for the
understanding of equilibrium solvation in axially-symmetric solvents:
(1) Solvation anisotropy, i.e.\ the dependence of the free energy of
solvation on the orientation of the solute dipole relative to nematic
director. (2) The effect of crossing the phase transition temperature
on the solvation thermodynamics. (3) The effect of dipolar
correlations on solvation and the applicability of continuum models of
solvation.

\begin{figure}[htbp]
  \centering
  \includegraphics*[width=8cm]{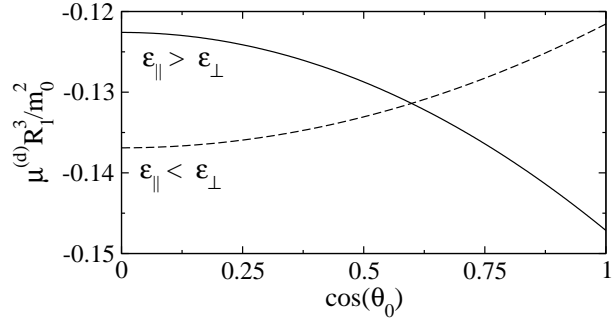}
  \caption{The solvation free energy of a dipole solute
    (in $m_0^2/R_1^3$ units) vs the angle $\theta_0$ between the solute
    dipolar moment and the nematic director (Fig.\ \ref{fig:1}); $R_0
    = 1.7 \sigma$.  Solid line refers to $\epsilon_{\parallel} > \epsilon_{\perp}$ (longitudinal
    solvent dipole, $\beta=0$) and the dashed line refers to $\epsilon_{\parallel} <
    \epsilon_{\perp}$ (transverse solvent dipole, $\beta=90$). Solvent parameters
    are: $\eta=0.47$, $S_2=0.8$, $y=0.15$, $\Lambda=0.3\sigma$. At $\beta=0$, the
    dielectric anisotropy is positive, $\epsilon_{\perp}=1.092$ and $\epsilon_{\parallel}=2.63$;
    at $\beta=90^0$, the dielectric anisotropy is negative, $\epsilon_{\perp}=1.746$
    and $\epsilon_{\parallel}=1.092$.  }
  \label{fig:6}
\end{figure}

From Eqs.\ (\ref{eq:2-36}), (\ref{eq:3-7}), and (\ref{eq:3-9}), the
solvation free energy of a dipole, $\mu^{(d)}$, is a quadratic function
of $\cos\theta_0$, where $\theta_0$ is the angle between the solute dipole and
nematic director (Fig.\ \ref{fig:1}).  Figure \ref{fig:6} shows that
the solvation anisotropy
\begin{equation}
  \label{eq:4-0}
  \Delta \mu^{(d)}=\mu^{(d)}_{\parallel}-\mu^{(d)}_{\perp}
\end{equation}
is negative in nematics with positive dielectric anisotropy
($\Delta\epsilon=\epsilon_{\parallel}-\epsilon_{\perp}>0$) and positive otherwise
($\mu^{(d)}_{\parallel}=\mu^{(d)}(\theta_0=0)$ and $\mu^{(d)}_{\perp}=\mu^{(d)}(\theta_0=\pi/2)$).
This can readily be verified by expanding Eq.\ (\ref{eq:2-36}) in
powers of the small parameter $\Delta\epsilon$:
\begin{equation}
\label{eq:4-1}
  \mu_{\text{c}}^{(d)}\approx-\frac{m_0^2}{2R_1^3}\left[ \frac{\epsilon_s-1}{2\epsilon_s+1} + \frac{2\Delta\epsilon}{5(2\epsilon_s+1)^2}P_2(\cos\theta_0) \right],
\end{equation}
where 
\begin{equation}
  \label{eq:4-1-1}
   \epsilon_s=(\epsilon_{\parallel}+2\epsilon_{\perp})/3.  
\end{equation}

In order to relate anisotropy of $\mu_{\infty}^{(d)}$ to dielectric
anisotropy, one needs a relation between the order parameter and the
dielectric constants. This connection is given by the Maier-Meier
theory:\cite{MaierMeier:ZNatur:61}
\begin{equation}
\label{eq:4-2}
\begin{split}
   \epsilon_{\parallel}-1 & = 3y\frac{3\epsilon_s}{2\epsilon_s+1}\left(1+2S_2P_2(\cos\beta)\right), \\
   \epsilon_{\perp}-1 & = 3y\frac{3\epsilon_s}{2\epsilon_s+1}\left(1-S_2P_2(\cos\beta)\right).
 \end{split}
\end{equation}
From Eq.\ \ref{eq:4-2}, $\Delta\epsilon\thicksim S_2P_2(\cos\beta)$. This means that
both $\Delta \mu_{\infty}^{(d)}$ and $\Delta \mu^{(d)}$ change their sign from negative
to positive when dielectric anisotropy changes its sign from positive
to negative.

The magnitude of solvation anisotropy $\Delta \mu^{(d)}$ is strongly affected by the size
of the solute. The continuum estimate of the solvation anisotropy parameter
\begin{equation}
  \label{eq:4-3}
  \kappa = \Delta \mu^{(d)} / \mu^{(d)}_{\text{av}},\quad \mu^{(d)}_{\text{av}} = \frac{1}{3}\left(\mu_{\parallel}^{(d)} + 2\mu_{\perp}^{(d)}\right)
\end{equation}
gives very low anisotropies ($\kappa\simeq0.02-0.03$) for a large number of 
nematics.\cite{LC:01} Once the dipole correlation effects are involved
through $\mu_{\infty}^{(d)}$, anisotropy becomes quite significant, $\kappa \simeq 0.3 S_2$. This
solvation anisotropy results in a discontinuity of the solvation chemical
potential at the point of the isotropic-nematic phase transition.
 
\begin{figure}[htbp]
  \centering
  \includegraphics*[width=6cm]{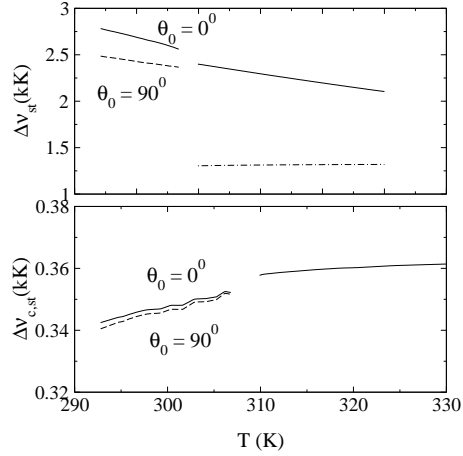}
  \caption{Stokes shift of coumarin-153 in nematic and
    isotropic phases of 5CB. The upper panel shows the microscopic
    calculation according to Eq.\ (\ref{eq:3-9}), the lower panel
    shows the continuum limit. The solid lines refer to the parallel
    alignment of the solute dipole with the nematic director, the
    dashed lines refer to the perpendicular alignment [Fig.\ 
    \ref{fig:1}]. The solute radius and the dipole moment are
    $R_0=4.89$ \AA{} and $\Delta m_0=7.5$ D, respectively. The temperature
    dependent dielectric constants,\cite{Horn:78,UrbanPCCP:99} order
    parameter,\cite{Horn:78} and density\cite{UrbanPCCP:99} of 5CB are
    taken from experiment; $\Lambda=0.35\sigma$. The dash-dotted line in the upper panel
    refers to the continuum isotropic result calculated with the
    cavity radius equal to $R_0$. }
  \label{fig:7}
\end{figure}

The significant effect of dipolar correlations on the solvation
thermodynamics is seen from comparison of lower and upper panels in
Fig.\ \ref{fig:7} which shows steady-state Stokes shift of
coumarin-153 dissolved in 4-n-pentyl-$4'$-cyanobiphenyl (5CB)
nematogen. This chromophore is widely used as a spectroscopic probe of
solvation dynamics and thermodynamics.\cite{Maroncelli:93,Reynolds:96}
Stokes shift dynamics of coumarin-153 in the isotropic phase of a
liquid crystalline solvent\cite{Rau:01} and of
coumarin-503\cite{Saielli:98,Bartolini:99} in both isotropic and
nematic phases have been reported.  The continuum limit in the lower
panel reveals a much weaker anisotropy in the nematic phase than the
full microscopic calculation in the upper panel. The continuum
calculations are also much lower in the absolute magnitude, which is
normally off-set by choosing the radius $R_0$ instead of the closest
approach distance $R_1$ (Fig.\ \ref{fig:7}, dash-dotted line).
However, for the present calculation, re-scaling the cavity radius
does not fully recover the solvation energy. This result suggests that
polar nematics might produce stronger solvation than isotropic
solvents with a comparable dielectric constant.

The electrostatic field of coumarin-153 is similar to that of a point
dipole,\cite{Kumar:95} which makes it a convenient system to test our
theory.  Since the Stokes shift experiments measure only nuclear
solvation, the Stokes shift $hc\Delta\bar \nu_{\text{st}}$ ($\bar \nu$ is the
wavenumber, cm$^{-1}$) was calculated according to the following
expression
\begin{equation}
  \label{eq:4-4}
    hc\Delta\bar \nu_{\text{st}} = - 2\mu_n^{(d)}(m_0=\Delta m_0) .
\end{equation}
Here, the difference in the dipole moments in the excited and ground
states of the chromophore $\Delta m_0$ is substituted for the solute dipole
moment. The nuclear component of solvation is calculated in the
additive approximation\cite{DMjpca:04} in which the overall solvation
free energy in a polar/polarizable liquid is assumed to be the sum of
the nuclear and electronic solvation components. The overall solvation
free energy $\mu(\epsilon_{\parallel},\epsilon_{\perp},y_{\text{eff}})$ is calculated from the
anisotropic dielectric constant in the component 
$\mu_c(\epsilon_{\parallel},\epsilon_{\perp})$ and the effective dipolar density
$y_{\text{eff}}$ in $\mu_{\infty}(y_{\text{eff}})$. The effective dipolar density is defined
as\cite{SPH:81}
\begin{equation}
  \label{eq:4-4-1}
  y_{\text{eff}} = (4\pi/9k_{\text{B}}T) \rho (m')^2 + (4\pi/3) \rho\alpha ,
\end{equation}
where $\alpha$ is the dipolar polarizability and $m'$ is the average dipole moment
of the solvent in the liquid. The nuclear free energy of solvation is then given by
\begin{equation}
  \label{eq:4-4-2}
  \mu_n = \mu(\epsilon_{\parallel},\epsilon_{\perp},y_{\text{eff}}) -  \mu(n_{\parallel}^2,n_{\perp}^2, y_e ) ,
\end{equation}
where $n_{\perp,\parallel}$ is the anisotropic refractive index and the density of
induced dipoles is
\begin{equation}
  \label{eq:4-4-3}
  y_e = (4\pi/3) \alpha\rho  .
\end{equation}
For coumarin-153, the radius $R_0=4.89$ \AA{} and the dipole moment change
$\Delta m_0=7.53$ D have been adopted.\cite{DMjpca:01}

5CB was chosen as a typical nematogen with its physical properties
well documented in the literature: the isotropic-nematic transition
temperature, $T_{\text{IN}}=308.2$ K\cite{Urban:99}, dipole moment,
$m=4.75$ D,\cite{UrbanPCCP:99} and the azimuthal angle between the
dipole moment and the long axis, $\beta=21.6^{0}$.\cite{Urban:00} The
temperature dependences of the static dielectric
constants,\cite{Urban:00,UrbanPCCP:99} refraction
indexes,\cite{Horn:78,UrbanPCCP:99} the order parameter,\cite{Horn:78}
and density\cite{UrbanPCCP:99} have also been reported.  For the
solvent diameter the value for benzene, $\sigma=5.27$ \AA,\cite{DMjpca:01}
was adopted, and the polarization correlation length was calculated
from Fig.\ \ref{fig:3} as $\Lambda=0.35\sigma$ corresponding to $(m^*)^2=3.3-3.6$.  The
dipole moment $m'$ was calculated using the Onsager approximation
\begin{equation}
  \label{eq:4-4-4}
   m' = \frac{(n^2+2)(2\epsilon_s + 1)}{3(2\epsilon_s + n^2)} m  , 
\end{equation}
where $\epsilon_s$ is given by Eq.\ (\ref{eq:4-1-1}) and $n^2 = (n^2_{\parallel} + 2
n^2_{\perp})/3$.

Note that this approximation is made in the Maier-Meier theory [Eq.\ 
(\ref{eq:4-2})] which, nevertheless, describes dielectric properties
of polar nematics reasonably well.\cite{Kresse:83,Bose:87} All results in
Fig.\ \ref{fig:7} have been obtained at experimentally documented
parameters of 5CB, the gap between the Stokes shift curves around the
isotropic-nematic transition temperature reflects the absence of
experimental data in this temperature range.

\subsection{Stokes shift dynamics}
\label{sec:4-2}
The dynamics of solvation following the photoinduced change in the
solute charge distribution is recorded by measuring the Stokes shift
correlation function\cite{Maroncelli:87}
\begin{equation}
\label{eq:4-6}
S(t)=\frac{E(t)-E(\infty)}{E(0)-E(\infty)},
\end{equation}
where $E(t)$ is the time-dependent energy of the solute.  The calculation
of this function is normally accomplished within the linear response
theory. Several formulations of the theory are available in the
literature,\cite{Wolynes:87,Fried:90,Bagchi:91} and we adopt here the
formulation due to Wolynes\cite{Wolynes:87} which represents the
Laplace transform $E(s)$ of the time-dependent function $E(t)$ as the
equilibrium solvation energy characterized by the dielectric constant
$\epsilon(s)$
\begin{equation}
\label{eq:4-7}
E(s)=E(\epsilon_{\parallel}(s),\epsilon_{\perp}(s))=
        \frac{2}{s}\left(\mu(\epsilon_{\parallel}(s),\epsilon_{\perp}(s),y_{\text{eff}})- \mu(n_{\parallel}^2,n_{\perp}^2,y_e)\right) .
\end{equation}

The dependence of $\Lambda$ on the dielectric constant of a nematic solvent
is generally unknown. We will therefore assume this parameter
independent of the Laplace variable $s$. Within this approximation, the
Stokes shift function is fully determined by the continuum expression
for the solvation energy [Eq.\ (\ref{eq:2-36})]. The results for
$S(t)$ presented here are therefore based on the continuum solution
for the dipolar solvation energy in an anisotropic dielectric with
axial symmetry.

Experimental dielectric data on nematics are well characterized
by the two-exponential form
\begin{equation}
\label{eq:4-8}
      \epsilon_a(s)=\epsilon_{\infty a}+(\epsilon_{0a}-\epsilon_{\infty a})\sum\limits_{k=1,2}\frac{g_k}{1+s\tau_{ka}} ,
\end{equation}
where $g_1+g_2=1$ and $a$ stands for $\parallel$, $\perp$ or $iso$ (isotropic).  This
form reflects two different relaxations: high-frequency rotation
around a long molecular axis (relaxation time $\tau_2$) and low-frequency
rotation around a short molecular axis (relaxation time
$\tau_1$).\cite{Urban:99} In isotropic phase, $\tau_{2iso}$ is about ten times
smaller that $\tau_{1iso}$, and the contribution from high-frequency
rotation to dielectric loss is usually small and is rarely resolved in
the dielectric experiments.\cite{LC:01,Urban:99} With
only one relaxation time $\tau_{1iso}$, Eq.\ (\ref{eq:4-8}) reduces to the
Debye dispersion resulting in a single-exponential Stokes shift
correlation function with the relaxation time $\tau_{S}=[(2\epsilon_{\infty} +
1)/(2\epsilon_{0} + 1)]\tau_{1iso}$ (Ref.\ \onlinecite{Bagchi:91}).
 
The liquid crystalline order of the nematic phase hinders rotations
around a short molecular axis when it is perpendicular to the nematic
director. The corresponding relaxation time $\tau_{1\parallel}$ is about ten
times larger than $\tau_{1iso}$. In contrary, $\tau_{1\perp}$ is much smaller
than $\tau_{1iso}$, and can be comparable to $\tau_{2a}$
($a=\parallel,iso$).\cite{LC:01}

With the dielectric constant from Eq.\ (\ref{eq:4-8}) substituted into
Eq.\ (\ref{eq:2-36}) the function $S(t)$ needs to be calculated
numerically. Following the procedure described in Ref.\ 
\onlinecite{DMjcp1:05}, the function $F(s) = -s\mathit{E}(s)$ was fitted
to a sum of Cole-Davidson type functions
\begin{equation}
 \label{eq:4-9}
      F(s)=\left(E(0)- E(\infty)\right) \sum\limits_{i} \frac{a_i}{(1+s\tau_{Si})^{\gamma_i}}
\end{equation}
with the exponents $\gamma_i$ and Stokes shift relaxation times $\tau_{Si}$;
the linear expansion coefficients are normalized by the condition
$\sum_ia_i=1$. Equation (\ref{eq:4-9}) allows analytical inverse Laplace
transform:
\begin{equation}
    \label{eq:4-10}
       S(t)=\sum_ia_i\frac{\Gamma(\gamma_i,t/\tau_{Si})}{\Gamma(\gamma_i,0)},
\end{equation}
where $\Gamma(\gamma,t)$ is the incomplete gamma function.\cite{GR}

\begin{figure}[htbp]
  \centering
  \includegraphics*[width=8cm]{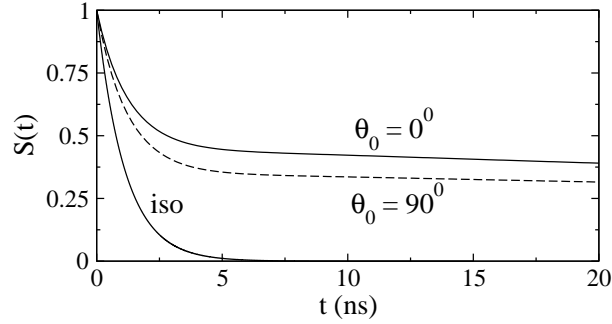}
  \caption{Stokes shift correlation function in the isotropic phase (''iso'')
    and in the nematic phase at two orientations of the solute relative
    to the nematic director.}
  \label{fig:8}
\end{figure}

Function $S(t)$ calculated for a spherical dipolar dye in
4,4-n-heptyl-cyanopiphenyl (7CB) is presented in Fig.\ \ref{fig:8}.
Experimental data for dielectric constants, refractive indexes and
low-frequency relaxation times is taken from Ref.\ 
\onlinecite{Davies:76}. No high-frequency relaxation times $\tau_{2a}$
have been reported for this nematogene.  $S(t)$ in the isotropic phase
(line marked ``iso'' in Fig.\ \ref{fig:8}) is single-exponential, as
expected.

\begin{figure}[htbp]
  \centering
  \includegraphics*[width=8cm]{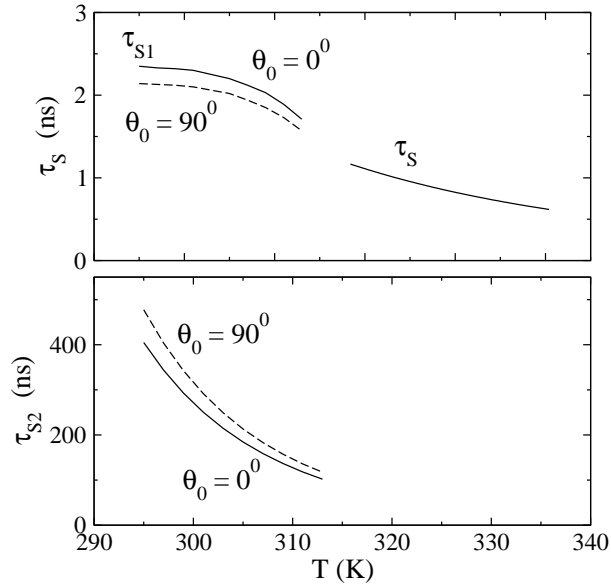}
  \caption{Temperature dependence of the Stokes shift relaxation times.  
    The upper panel shows the faster relaxation time in the isotropic
    ($\tau_S$) and nematic ($\tau_{S2}$) phases. The lower panel shows the
    slower relaxation time ($\tau_{S1}$) present only below $T_{\text{IN}}$. The
    solid lines refer to the orientation of the solute dipole parallel
    to the nematic director, while the dashed lines refer to the
    perpendicular orientation. }
  \label{fig:9}
\end{figure}

Two predictions follow from our calculations in the nematic phase.
First, $S(t)$ is bi-exponential.  The slow component with
$\tau_{S1}\approx100-500$ ns and $\gamma_{1}\approx0.82$ is related to $\epsilon_{\parallel}$, and the
fast component with parameters $\tau_{S2}\approx1-2$ ns and $\gamma_{2}\approx1.2$ is
related to the relaxation of $\epsilon_{\perp}$.  Second, $S(t)$ is effected by
the angle $\theta_0$ between the solute dipole and the nematic director.
For nematics with positive dielectric anisotropy (e.g., 7CB) $S(t)$
decays faster for $\theta_0=90^0$ than for $\theta_0=0^0$. Note that the slow
component has not been detected in experimental studies of
transient Stokes shift in
nematics,\cite{Bartolini:99,Saielli:98,Rau:01} probably because of the
limited experimental time resolution (a few ns). All Stokes shift
relaxation times monotonically increase with lowering temperature 
in both the nematic and isotropic phase (Fig.\ \ref{fig:9}).

\section{Conclusions}
\label{sec:5}
This article presents a microscopic theory of solvation in solvents
with axial symmetry. Although applications of the theory considered
here are limited to nematic liquid crystals, i.e.\ liquids with
inversion symmetry of the polar axis, the formalism is also applicable
to ferroelectric solvents with a preferential polar direction.  This
conclusion follows from the fact that the linear response
approximation leads to a quadratic dependence on the solute dipole
moment invariant to the dipole flip.  For ferroelectrics, however, the
solvation energy gains an additional contribution, linear in the
solute dipole, from the macroscopic polarization of the solvent. 

The full microscopic formulation requires $k$-dependent dipolar
susceptibility of the nematic solvent which needs to be obtained from
computer experiment. From the analysis of the results of MC
simulations and the microscopic formalism, we have derived a formula
for the solvation chemical potential which is based on experimentally
measurable input parameters and a theory parameter, the length of
dipolar correlations in the solvent. This correlation length was
obtained from MC simulations in the range of parameters attainable for
the model fluid of dipolar spherocylinders.  The understanding of the
properties of the correlation length in a broader range of parameters,
in particular for higher dielectric constants, will require
simulations of nematogenes with more realistic intermolecular
potentials.

The theory provides a quantitative framework for interpreting the
spectroscopic steady-state and time-resolved experiments and makes
several experimnetally testable predictions. We show that the
equilibrium free energy in the nematic phase is a quadratic function
of cosine of the angle between the solute dipole and the solvent
nematic director. The sign of solvation anisotropy is determined by
the sign of dielectric anisotropy of the solvent: solvation anisotropy
is negative in solvents with positive dielectric anisotropy and
\textit{vice versa}.  The solvation free energy is discontinuous at
the point of isotropic-nematic transition.  The amplitude of this
discontinuity is strongly affected by the size of the solute becoming
less pronounced for larger solutes.  The discontinuity itself and the
magnitude of the splitting of the solvation chemical potential in the
nematic phase are mostly affected by microscopic dipolar correlations
in the nematic solvent.

The Stokes shift correlation function in the isotropic phase is
one-exponential when dielectric relaxation is given by the Debye form.
The corresponding relaxation time is a smooth function of temperature
through isotropic-nematic transition. In the nematic phase, the Stokes
shift relaxation becomes bi-exponential with a much slower relaxation
component related to rotations of polar molecules around their short
axes in the nematic potential strongly hindering such motions.

\appendix

\section{Derivation of Eq.\ (\ref{eq:2-26})}
\label{appendix}
We start with the formal expression for the solvent dipolar susceptibility\cite{Klapp:00}:
\begin{equation}
\begin{split}
\pmb{\tilde \chi}_s(\mathbf{k}) = & (m^2/k_{\text{B}}T) \int d\omega_1 d\omega_2 \\
 &\mathbf{\hat m}(\omega_1) 
\mathbf{\hat m}(\omega_2) \left[ \delta(\omega_1-\omega_2) \rho(\omega_1) + \rho(\omega_1) 
\rho(\omega_2) \tilde{h}(\mathbf{k},\omega_1,\omega_2) \right],
\end{split}
\end{equation}
where $m$ and $\omega$ are the magnitude and orientation of the solvent dipole. Further, 
the pair correlation function $\tilde{h}(\mathbf{k},\omega_1,\omega_2)$ is expanded in spherical
harmonics\cite{Gubbins:84,Klapp:00}
\begin{equation}
\label{A3}
\tilde{h}(\mathbf{k},\omega_1,\omega_2)= \sum\limits_{l_1 l_2 l} \sum\limits_{n_1 n_2} \tilde{h}_{n_1 n_2}^{l_1 l_2 l}(k) 
Y_{l_1,n_1}(\omega_1) Y_{l_2,n_2} (\omega_2) Y^*_{l,n_1+n_2}(\omega_k),
\end{equation}
where $\tilde{h}_{n_1 n_2}^{l_1 l_2 l}(k)$ is the Hankel transform:
\begin{equation}
\label{A2}
\tilde{h}_{n_1 n_2}^{l_1 l_2 l}(k) = 4\pi i^l \int\limits_0^{\infty} dr\, r^2\, j_l(kr)
h_{n_1 n_2}^{l_1 l_2 l}(r) .
\end{equation}
Then the tensor $\pmb{\tilde \chi}_s$ can be written as:
\begin{equation}
\label{A4}
\tilde{\chi}_{s, n_1 n_2}(\mathbf{k}) = \sum\limits_l \tilde{\chi}_{s,n_1 n_2 l}(k)
Y^*_{l,-n_1-n_2}(\omega_k)
\end{equation}
where coefficients $\tilde{\chi}_{s,n_1 n_2 l}(k)$ are proportional to
$\tilde{h}_{n_1 n_2}^{l_1 l_2 l}(k)$ and depend only on the magnitude
of wave-vector $k$.

We next prove the relation
\begin{equation}
\label{A5}
\pmb{\tilde \chi}_s(k=0) - \pmb{\tilde \chi}'(k=0) = \int \frac{d\omega_k}{4\pi} \pmb{\tilde \chi}_s(k=0)
\end{equation}
where $\pmb{\tilde \chi}'$ is the part of $\pmb{\tilde \chi}_s$ which arises
from solvent occupied the volume outside of the solute. Since the
function $\pmb{\tilde \chi}'$ is defined by integrating over the volume
twice larger than the solute volume it depends only on the asymptote
of solvent correlation function on long distances. We will seek the
asymptotes of $h^{l_1 l_2 l}_{n_1 n_2}(r)$ in form $1/r^n$. Since
$h(12)$ cannot decay slower than the interaction potential, $n\geq3$.
According to (\ref{A4}) and (\ref{A2}) the contribution from the
asymptotes is proportional to
\begin{equation}
I=k^{n-3} \int\limits_{kd}^{\infty} dx \frac{j_l(x)}{x^{n-2}}
\end{equation}
where $d$ is an arbitrary length larger than size of solvent particle.

We first consider the case $n=3$:
\begin{equation}
I=\int\limits_{kd}^{\infty} dx \frac{j_l(x)}{x}
\end{equation}
In the limit $k\to0$, this integral converges only if $l>0$. Since
$\pmb{\tilde \chi}_{s}(\mathbf{k})$ is analytical function at $k=0$, only
harmonics with $l>0$ can have asymptote $1/r^3$.  In case of $n>3$ we
split $I$ into two parts:
\begin{equation}
I=k^{n-3} \int\limits_{\delta}^{\infty} dx \frac{j_l(x)}{x^{n-2}} + 
k^{n-3} \int\limits_{kd}^{\delta} dx \frac{j_l(x)}{x^{n-2}},
\end{equation}
where parameter $\delta$ is small enough that $j_l(x)$ can be replaced by
the first term of its Taylor expansion $x^l/(2l+1)!!$. Then the first
part of $I$ vanishes in limit $k\to0$.  The second part is proportional
to
\begin{equation}
k^{n-3}\int\limits_{kd}^{\delta} \frac{x^{l-n+2}}{(2l+1)!!} \thicksim k^l
\end{equation}
and in the limit $k\to0$ gives non-zero contribution only into term with
$l=0$. Therefore, in long-wave limit term $l=0$ in Eq.\ (\ref{A4})
arises from harmonics with asymptotes $1/r^n$ where $n>3$, and all
therms $l>0$ steam from harmonics with asymptotes $1/r^3$.  Assuming
that only the $1/r^3$ asymptotes contribute to $\tilde{\chi}'(k=0)$, we
arrive in Eq.\ (\ref{A5}) from Eq.\ (\ref{A4}) and orthogonality of
spherical harmonics.

\begin{acknowledgments}
  This research was supported by the National Science Foundation
  (CHE-0304694). This is publication \#646 from the ASU Photosynthesis
  Center.
\end{acknowledgments}

\bibliographystyle{apsrev}
\bibliography{kapko_asu,/home/dmitry/p/bib/chem_abbr,/home/dmitry/p/bib/photosynth,/home/dmitry/p/bib/et,/home/dmitry/p/bib/liquids,/home/dmitry/p/bib/solvation,/home/dmitry/p/bib/dynamics,/home/dmitry/p/bib/dm,/home/dmitry/p/bib/lc,/home/dmitry/p/bib/lcold}

\end{document}